\documentclass{ws-procs975x65}
\usepackage{indentfirst}

\begin{document}

\title{\uppercase{On the stability of circular orbits in galactic dynamics:
Newtonian Thin Disks}}

\author{\uppercase{Ronaldo S. S. Vieira}$^*$}

\address{Instituto de F\'isica ``Gleb Wataghin'', Universidade Estadual de Campinas,\\
Campinas, SP 13083-859, Brazil\\
$^*$E-mail: ronssv@ifi.unicamp.br}

\author{\uppercase{Javier Ramos-Caro}}

\address{Departamento de F\'{i}sica, Universidade Federal de S\~{a}o Carlos,\\
S\~{a}o Carlos, SP 13565-905, Brazil}

\begin{abstract}
The study of off-equatorial orbits in razor-thin disks is still in its beginnings. Contrary to what was presented in the
literature in recent publications, the vertical stability criterion for equatorial circular orbits cannot be based on the
vertical epicyclic frequency, because of the discontinuity in the gravitational field on the equatorial plane.
We present a rigorous criterion for the vertical stability of circular orbits in systems composed by a razor-thin disk surrounded
by a smooth axially symmetric distribution of matter, the latter representing additional structures such as thick disk, bulge
and (dark matter) halo. This criterion is satisfied once the mass surface density of the thin
disk is positive. Qualitative and quantitative analyses of nearly equatorial orbits are presented. In particular, the
analysis of nearly equatorial orbits allows us to construct an approximate analytical third integral of motion in this region of
phase-space, which describes the shape of these orbits in the meridional plane.
\end{abstract}

\keywords{Galactic dynamics; Theory of orbits}

\bodymatter

\bodymatter\bigskip
%\section*{}
%\section{Introduction}

\

The literature on analytical models for disklike galaxies is vast (see, for instance, Ref.~\refcite{BT2008}).
Although most disk galaxies
contain internal structures such as bars, warps and spiral arms \cite{vanderkruit2011},
we can gain a deep insight about the orbital dynamics of stars in these systems by modeling them as axisymmetric. This is a
standard procedure to model, for instance, rotation curves of spiral galaxies. In this framework, a given rotation curve
is usually fitted by an axisymmetric mass model decomposition into a stellar disk, a gaseous disk, a central bulge and a
dark matter halo. By assuming a constant mass-to-light ratio along the galaxy, it is possible to determine
the density profiles of the disks by photometry\cite{vanderkruit2011}.

There is one extra aspect which must be considered in modeling rotation curves of disk galaxies: the corresponding mass models
must give rise to stable circular orbits in the disk's equatorial plane. If we consider a smooth
3D distribution of matter, the stability
analysis follows from Taylor expanding the effective potential up to second order in the vincinity of the circular orbit
\cite{BT2008}, which is stable if the epicyclic frequencies $\kappa$ (radial) and $\nu$ (vertical) are real. However, many
galaxy models in the literature are composed by a razor-thin disk. Although in this case radial stability
of circular orbits is analyzed as above, 
vertical stability cannot be analyzed by second-order Taylor expansion. This happens because of the discontinuity in 
the vertical component of the gravitational field, due to the surface layer of matter in the equatorial plane. 
In this contribution we briefly present
the results of our recent work \cite{vieira12} concerning the vertical stability of circular orbits in Newtonian
razor-thin disks. After presenting the basics of razor-thin disk modeling, we comment on the new vertical stability criterion,
which is consistent with the delta-like singularity in the density field. Adiabatic approximation permits us to construct an
approximate third integral of motion for nearly equatorial orbits in these systems, which describes the shape of
the corresponding
3D orbits in the $Rz$ plane. In this way, we expect that the formalism presented here can also contribute to the
``third integral of motion'' issue \cite{hunter05}, as well as to the analysis of orbits in piecewise-smooth
dynamical systems. Concerning this last topic, we remark that, as far as we know, there is no analogue of the KAM theorem
which is valid for systems with this kind of discontinuity. However, the literature on disk-crossing orbits presents
a large amount of numerical evidence in which islands of stability appear in the vicinity of a 
stable fixed point
\cite{hunter05, ramos-caro08, ramos-caro11}.

The density profile is described as
  \begin{equation}
   \rho(R,z) = \Sigma(R)\delta(z) + \rho_s(R,z),
  \end{equation}
where $\delta(z)$ is the Dirac delta distribution, $\rho_s$ is smooth and the system has reflection symmetry with respect to
$z=0$. The physical interpretetion of the above quantities is the following:
$\Sigma(R)$ is the surface mass density profile of the razor-thin disk and $\rho_s(R,z)$ represents all the remaining
three-dimensional structures, such as thick disk, bulge and (dark matter) halo.
For an equatorial circular orbit of radius $R_o$, we found \cite{vieira12} that the condition which must substitute the vertical
frequency stability condition is
  \begin{equation}\label{sigma>0}
   \Sigma(R_o)>0.
  \end{equation}
The radial stability condition is the same as in 3D disks, i.e., $\kappa^2(R_0)>0$. Moreover, $\Sigma>0$ and $\kappa^2>0$
 imply (Liapunov) stability of the corresponding circular orbit \cite{vieira12}. The vertical stability condition
presented here is satisfied along the whole region of the equatorial plane which contains the surface layer of matter. Therefore,
every equatorial circular orbit in an axially symmetric razor-thin disk model 
is vertically stable by construction, regardless of the shape of the remaining components of the system.

The procedure considered so far in the literature to analyze vertical stability of circular orbits in analytical models of
razor-thin disks took into account the vertical epicyclic frequency criterion \cite{ramos-caro08, gonzalez10, ramos-caro11},
neglecting the term proportional to $\delta(z)$. This led to the conclusion that some models contain regions with
vertically unstable circular orbits\cite{ramos-caro08, gonzalez10, ramos-caro11}. A suggested solution to this problem
was to consider, in addition to the razor-thin disk, thick disks or spheroidal halos in order to guarantee vertical stability
\cite{gonzalez10}. However, as described in the preceding paragraph (and discussed in detail in Ref.~\refcite{vieira12}),
this is not the case. The addition of 3D distributions of matter does not affect vertical stability in
such systems. Radial stability, though, is affected by the off-equatorial density profile $\rho_s$.

\begin{figure}[h]
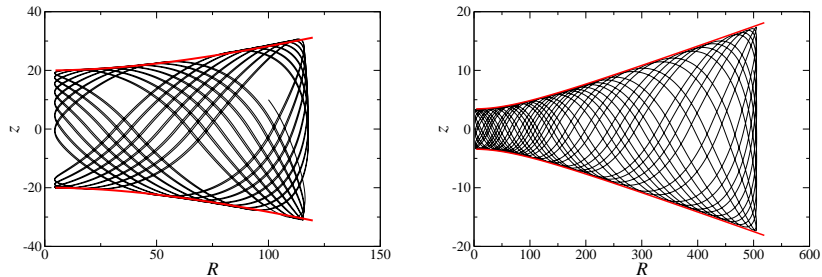
%
\begin{center}
 \parbox{2.1in}{\epsfig{figure=k1.eps,width=2in}}
 \hspace*{4pt}
 \parbox{2.1in}{\epsfig{figure=k2.eps,width=2in}}
 \caption{Orbits in Kuzmin's potential. The solid (red) line corresponds to the prediction of (\ref{AA}).
 More details can be found in Ref.~\refcite{vieira12}.}
\label{fig:k}
\end{center}
\end{figure}

Concerning nearly equatorial orbits close to a stable equatorial circular orbit with given $z$ component of angular momentum,
adiabatic invariance of the approximate Hamiltonian describing vertical motion near the circular orbit
(with $R(t)$ acting as an external time dependence in the corresponding potential)
gives us the following relation \cite{vieira12} for the shape of nearly equatorial orbits in the meridional plane:
  \begin{equation}\label{AA}
   \frac{Z(R)}{Z(R')} = \bigg(\frac{\Sigma(R')}{\Sigma(R)}\bigg)^{1/3},
  \end{equation}
where $Z(R)$ is the maximum value of $z$ that the star reaches along the whole trajectory, for each given value of $R$.
This expression
represents an approximate third integral of motion for the system in the phase-space region under consideration, and depends
only on the density distribution of the razor-thin disk. The validity of this relation was checked numerically in
Ref.~\refcite{vieira12} for a Kuzmin thin disk and a Kuzmin disk superposed by a Plummer halo, with good agreement near the
equatorial plane (see Fig. \ref{fig:k} for orbits in Kuzmin's potential). This approximate third integral of motion may be
a starting point to apply canonical perturbation theory to integrable Hamiltonian systems containing razor-thin disks.

\section*{Acknowledgements}

The authors thank FAPESP for financial support.


\begin{thebibliography}{}

\bibitem{BT2008}
Binney, J., Tremaine, S., \textit{Galactic Dynamics 2nd. edition}, Princeton University Press (2008).

\bibitem{vanderkruit2011}
van der Kruit, P. C., Freeman, K. C., {\em ARAA} {\bf 49}, 301 (2011).

\bibitem{vieira12}
Vieira, R. S. S., Ramos-Caro, J., arXiv:1206.6501 [astro-ph.CO].

\bibitem{hunter05}
Hunter, C., {\it Ann.\ NY.\ Acad.\ Sci.} {\bf 1045}, 120-138 (2005).

\bibitem{ramos-caro08}
Ramos-Caro, J., L\'opez-Suspes, F., Gonz\'alez, G. A., {\em MNRAS} {\bf 386}, 440 (2008).

\bibitem{ramos-caro11}
Ramos-Caro, J., Pedraza, J. F., Letelier, P. S., {\em MNRAS} {\bf 414}, 3105-3116 (2011).

\bibitem{gonzalez10}
Gonz\'alez. G. A., Plata-Plata, S. M., Ramos-Caro, J., {\em MNRAS} {\bf 404}, 468-474 (2010).



\end{thebibliography}
\end{document}